\documentclass[english,twocolumn,showpacs,preprintnumbers,amsmath,amssymb,superscriptaddress,prl]{revtex4}
\usepackage{mathptmx}
\usepackage[T1]{fontenc}
\usepackage[latin9]{inputenc}
\usepackage{units}
\usepackage{amsmath}
\usepackage{amssymb}
\usepackage{graphicx}
\usepackage{esint}

\makeatletter
\@ifundefined{textcolor}{}
{%
 \definecolor{BLACK}{gray}{0}
 \definecolor{WHITE}{gray}{1}
 \definecolor{RED}{rgb}{1,0,0}
 \definecolor{GREEN}{rgb}{0,1,0}
 \definecolor{BLUE}{rgb}{0,0,1}
 \definecolor{CYAN}{cmyk}{1,0,0,0}
 \definecolor{MAGENTA}{cmyk}{0,1,0,0}
 \definecolor{YELLOW}{cmyk}{0,0,1,0}
}

\makeatother

\usepackage{babel}
\begin{document}

\title{Unveiling the orbital angular momentum and acceleration of electron
beams}

\author{Roy Shiloh}

\email{royshilo@post.tau.ac.il}

\affiliation{Department of Physical Electronics, Fleischman Faculty of Engineering,
Tel Aviv University, Tel Aviv 6997801, Israel}

\author{Yuval Tsur}

\affiliation{Department of Physical Electronics, Fleischman Faculty of Engineering,
Tel Aviv University, Tel Aviv 6997801, Israel}

\author{Yossi Lereah}

\affiliation{Department of Physical Electronics, Fleischman Faculty of Engineering,
Tel Aviv University, Tel Aviv 6997801, Israel}

\author{Boris A. Malomed}

\affiliation{Department of Physical Electronics, Fleischman Faculty of Engineering,
Tel Aviv University, Tel Aviv 6997801, Israel}

\author{Vladlen Shvedov}

\affiliation{Laser Physics Centre, The Australian National University, Canberra
ACT 0200, Australia}

\author{Cyril Hnatovsky}

\affiliation{Laser Physics Centre, The Australian National University, Canberra
ACT 0200, Australia}

\author{Wieslaw Krolikowski}

\affiliation{Laser Physics Centre, The Australian National University, Canberra
ACT 0200, Australia}

\author{Ady Arie}

\affiliation{Department of Physical Electronics, Fleischman Faculty of Engineering,
Tel Aviv University, Tel Aviv 6997801, Israel}
\begin{abstract}
New forms of electron beams have been intensively investigated recently,
including vortex beams carrying orbital angular momentum, as well
as Airy beams propagating along a parabolic trajectory. Their traits
may be harnessed for applications in materials science, electron microscopy
and interferometry, and so it is important to measure their properties
with ease. Here we show how one may immediately quantify these beams\textquoteright{}
parameters without need for additional fabrication or non-standard
microscopic tools. Our experimental results are backed by numerical
simulations and analytic derivation.
\end{abstract}
\maketitle
In the last few years it became possible to generate special shapes
of electron beams. One of these special beams is the vortex beam \cite{Verbeeck2010,McMorran2011,Uchida2010}
having a helical wavefront structure and a phase singularity on axis.
Its azimuthal phase dependence is $\exp\left(il\phi\right)$, where
integer $l$ is the topological charge and $\phi$ is the azimuthal
angle. This beam carries an orbital angular momentum of $\hbar l$
\cite{Allen1992}. Another interesting beam is the Airy beam, having
a transverse amplitude dependence in the form of the Airy function,
i.e. $\mathrm{Ai}\left(x/x_{0}\right)$, where $x_{0}$ defines the
transverse scale. It is shape-preserving and moves along a parabolic
trajectory in free-space with a nodal trajectory coefficient, sometimes
referred to as the ``acceleration'' coefficient, $1/2x_{0}^{3}k^{2}$,
where $k$ is the wave number. These beams, generated and observed
in a TEM (Transmission Electron Microscope) are expected to open new
possibilities for interactions between electrons and matter, as well
as for electron microscopy and interferometry. For example, vortex
beams were used in electron energy loss spectroscopy in order to characterize
the magnetic state of a ferromagnetic material \cite{Verbeeck2010},
while Airy beams were proposed for realization of a new type of electron
interferometer \cite{Voloch-Bloch2013}. In order to utilize these
new types of electron beams, it is required to develop methods that
allow one to easily determine their defining properties - the OAM
in the case of a vortex beam or the nodal trajectory coefficient in
the case of an Airy beam. 

Recently, a new method was proposed for measuring the OAM of electron
beams \cite{Saitoh2013} in TEM. Two holographic plates were used:
a vortex-generating spiral plate in the condenser aperture and a fork
grating acting as an analyzer in the selected-area aperture. This
method relies on a custom modification of the objective aperture,
which requires technical specialists and is less desired in routinely-operated
facilities, though it is advantageous for mixed-state OAM-carrying
modes. In this Letter, we demonstrate a straightforward method for
OAM pure-state determination, which dispenses with fabrication or
aperture-manipulation. We then extend the idea to the measurement
of the nodal trajectory of electron Airy beams, along with supporting
simulations and mathematical derivation.

In their 1991 article \cite{Abramochkin1991} Abramochkin and Volostnikov
discussed the mathematics of beam transformations under astigmatic
conditions. Since then, different authors proposed and demonstrated
mode conversion in lasers \cite{Machavariani2004}, linear and nonlinear
optics \cite{Ellenbogen2008,Bloch2012} and free-electron beams in
TEM \cite{Schattschneider2012}. In the latter work, Schattschneider
et. al. have shown in theory and experiment how first-order Laguerre-Gauss
modes may be transformed to first-order Hermite-Gauss modes. Recently,
it was proposed that a cylindrical lens acting as a mode converter
\cite{Shvedov2010} be used to quantify the OAM of optical vortices
\cite{Denisenko2009}. Thus, a vortex of integer topological charge
converts into a corresponding Hermite-Gauss-like mode, where the number
of dark stripes precisely indicates the topological charge. The method's
advantage is in its simplicity and generality: the only addition to
the setup is a cylindrical lens. In the TEM, the lens is inherent
- the condenser and objective stigmators may be used to impose a strong
astigmatism along a desired axis, thereby implementing the elliptic
transformation required to perform mode conversion. 

Other OAM-measuring methods have previously been investigated \cite{Guzzinati2014},
some examples of which include interaction with additional holographic
plates \cite{Mair2001}, performing a geometric optical transformation
\cite{Lavery2011}, modal decomposition \cite{Schulze2013} and optical
transformations \cite{Mirhosseini2013}.

In our experiment, we designed a two-dimensional binary hologram -
a fork grating which is mathematically written

\begin{equation}
h\left(x,y\right)=\mathrm{sign}\left\{ \sin\left[\frac{2\pi x}{\Lambda}-l\cdot\mathrm{atan2}\left(y,x\right)+\delta\frac{\pi}{2}\right]+\Delta\right\} \label{eq:Fork}
\end{equation}
where $2\pi x/\Lambda$ is a linear grating along $x$, with period
$\Lambda$, upon which the four-quadrant inverse-tangent function%
\footnote{$\mathrm{atan2}(y,x)$ returns the angle in the interval $[-\pi,\pi]$
as opposed to $\mathrm{atan}$ which is limited to $[-\pi/2,\pi/2]$,
i.e. only the two quadrants in the positive $x$ half-space.%
} modulates the spiral phase of the vortex, $l$ being the topological
charge. $\delta$ reduces fabrication errors in the centre by preserving
continuity: thus its value is chosen equal to zero (one) if $l$ is
odd (even). $\Delta$ relates to duty cycle; we chose $\Delta=0.25$
so both even and odd diffraction orders are visible. A Raith IonLine
focused ion-beam (FIB) was used to mill into a $\unit[100]{nm}$ SiN
membrane coated by $\unit[50]{nm}$ Au, from the gold side. The function
of the gold layer is two-fold: to prevent charging of the sample and
to assure the membrane behaves as an amplitude (binary) grating.

\begin{figure}
\begin{centering}
\includegraphics[clip,width=1\columnwidth]{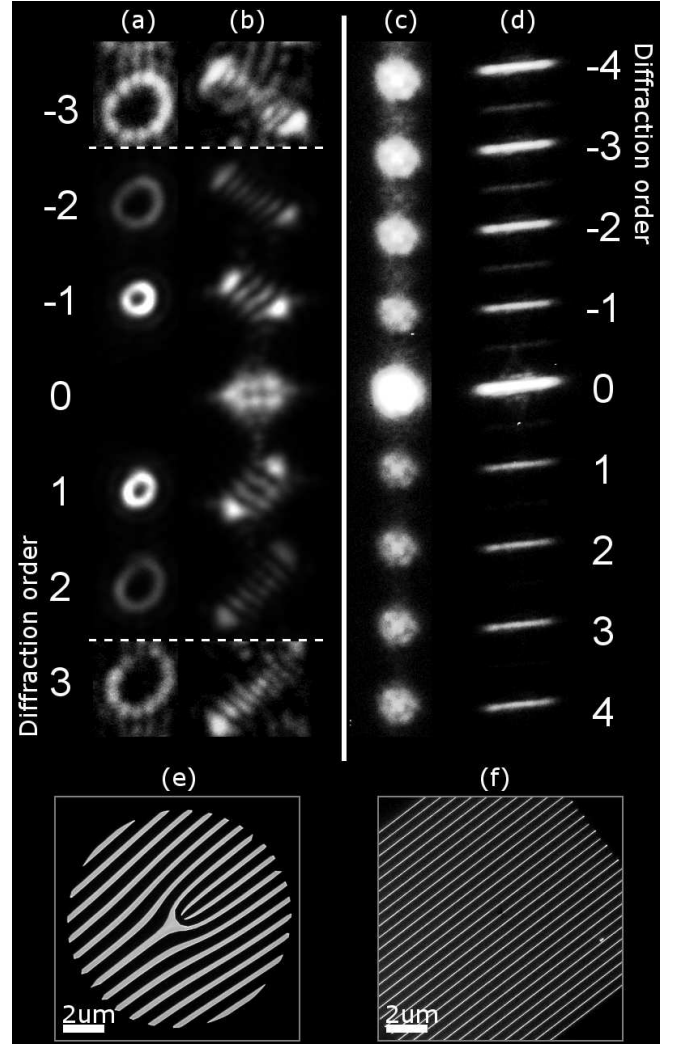}
\par\end{centering}

\centering{}\caption{\label{fig:charge3} (Color online) Experimental results: diffraction
patterns of (a) vortices generated by a charge 3 fork grating, using
a focused stigmatic beam (zero-order blocked); (b) corresponding vortices
after elliptic transformation; (c) diffraction pattern generated by
a linear calibration grating using a stigmatic beam, out of focus;
(d) corresponding lines resulting from the elliptic transformation,
out of focus. Below: TEM images of (e) charge 3 fork ($\unit[0.750]{\mu m}$
period); (f) linear grating ($\unit[0.445]{\mu m}$ period). Note:
brightness levels in the third order parts, as marked by the dashed
lines in (a) and (b), have been modified for visibility.}
\end{figure}

A fork grating with $l=3$ topological charge was fabricated (Fig.
\ref{fig:charge3}e), based on a $\Lambda=\unit[0.750]{\mu m}$ period,
along with a linear (Bragg) grating of period $\Lambda=\unit[0.445]{\mu m}$
which was fabricated for calibration purposes (Fig. \ref{fig:charge3}f).
The sample was mounted onto a Tecnai F-20 FEG-TEM with a single tilt
specimen holder, and subsequently observed in the microscope in Low
Angle Diffraction (LAD) mode through a $\unit[10]{\mu m}$ aperture. 

In order to form the far field diffraction pattern, the condenser
lens was set so a focused spot is measured. The fork grating was then
aligned under the beam and the hologram imposed a spiral phase such
that in each diffraction order $m$, a vortex of OAM $l\cdot m$ emerged
(Fig. \ref{fig:charge3}a). The intense on-axis (zero-order) beam
was blocked to prevent damage to the CCD.

Since the electron beam impinging on the hologram does not have a
Gaussian distribution, the emerging vortices are not strictly Laguerre-Gauss
modes. However, the resulting diffraction pattern is undeniably dominated
by solutions of the Schrödinger equation
\begin{equation}
\left[\nabla_{\bot}^{2}+2ik_{DB}\frac{\partial}{\partial z}\right]\psi=0\label{eq:Helmholtz}
\end{equation}

where $k_{DB}=1/\lambda_{DB}$ is the de-Broglie wave-number. This
equation governs the paraxial dynamics of the slowly varying envelope
of the free electron beam in the TEM\cite{reimer2008transmission},
a statement which is augmented by the results of the next step in
our experiment.

The electron beam is made sufficiently astigmatic along the desired
axis so it becomes elliptic; in the present case, this axis must be
parallel to the lateral direction of the fork. As a reference, the
linear grating was first measured under these conditions. As will
be explained below, we purposely measured slightly out of the focal
plane (Fig. \ref{fig:charge3}c). This reference serves to assess
the power of the astigmatism, or, the efficacy of the elliptic transformation.
When the beam is sufficiently elliptic, the vortices in the different
diffraction orders are converted into Hermite-Gauss-like modes (Fig.
\ref{fig:charge3}b) and the resulting dark stripes indicate the OAM
carried by each order. In addition, the angle of the transformed orders
signifies the rotation direction of the vortex: clockwise or counter-clockwise.
The linear grating's diffraction spots become lines, as expected from
a cylindrical lens (Fig. \ref{fig:charge3}d). Note that if the linear
grating's diffraction pattern was observed in focus rather than out
of it, the spots would be so small that the equipped Gatan 694 CCD
would not be able to distinguish the ellipticity of the beam; also,
the additional diffraction lines visible in the negative orders (Fig.
\ref{fig:charge3}d) are a result of asymmetry of the sample relative
to the beam, introduced by using a single-tilt rather than a double-tilt
holder. In the supplementary material, we show a similar, additional
measurement of at least $10\hbar$.

We then used the same principle to investigate the acceleration of
2D Airy beams. Optical generation and manipulation of Airy beams \cite{Berry1979,Siviloglou_OL_32_2007}
has been recently under the spot-light; here we present, to the best
of our knowledge, the first encounter of Airy beams with special transformations,
specifically the astigmatic transformation.

For the purpose of generality, we express the transformation of a
beam, $f\left(\xi,\eta\right)$, under general astigmatic conditions
by

\begin{equation}
F\left(k_{\xi},k_{\eta},a\right)=\int_{-\infty}^{\infty}\int_{-\infty}^{\infty}e^{i\left(k_{\xi}\xi+k_{\eta}\eta\right)+i\psi\left(\xi,\eta,a,\alpha\right)}f\left(\xi,\eta\right)d\xi d\eta.
\end{equation}

Here we define the Cartesian coordinates $\left(\xi,\eta\right)$
in the plane of the holographic mask, and similarly $\left(k_{\xi},k_{\eta}\right)$
in the diffraction plane. $\psi\left(\xi,\eta,a,\alpha\right)=a\left[\left(\xi^{2}-\eta^{2}\right)\mathrm{cos}2\alpha+2\xi\eta\mathrm{sin}2\alpha\right]$
and $a$,$\alpha$ are defined as in \cite{Abramochkin1991}. In that
paper, it is shown that in the special case of $\psi\left(\xi,\eta,a,\alpha=\pi/4\right)=2a\xi\eta$,
Hermite-Gauss and Laguerre-Gauss beams interchange; this mathematical
paradigm explains our previous results.

Applying the transformation to Airy beams, one may generally write

\begin{equation}
F\left(k_{\xi},k_{\eta},a\right)=\int_{-\infty}^{\infty}\int_{-\infty}^{\infty}e^{i\left(k_{\xi}\xi+k_{\eta}\eta\right)+2ai\xi\eta}e^{i\left(\xi^{3}+\eta^{3}\right)/3\widetilde{\beta}^{3}}d\xi d\eta,\label{eq:eq4_F(k_xi,k_eta,a)}
\end{equation}

where $\widetilde{\beta}$ relates to the beam's nodal trajectory
(sometimes referred to as ``acceleration'' coefficient, through
$\widetilde{\beta}^{3}/2$). In a diffractive system with focal length
$f$, we define 

\begin{equation}
\mathfrak{a}=\beta^{3}k_{\mathrm{DB}}/2f^{3},\label{eq:acceleration}
\end{equation}

where $\beta$ is defined below through fabrication parameters and
$k_{\mathrm{DB}}$ is the de-Broglie wavelength. While calculation
of this integral in the stigmatic ($a=0$) case is easy and results
in the two-dimensional Airy pattern, obtaining a closed-form analytic
solution to the astigmatic ($a\neq0$) case is difficult. In order
to measure the nodal trajectory coefficient from a stigmatic pattern,
one must either record the propagation of the Airy beam in different
planes and measure the actual trajectory, or directly measure the
density of the Airy lobes and relate it to the length scale $x_{0}$
(for more details, see \cite{Voloch-Bloch2013}). These two methods
are, unfortunately, time consuming and arduous; the latter case, for
example, is generally dangerous to perform experimentally, since such
a measurement requires the pattern to be in focus, to exhibit high-contrast
and low signal-to-noise ratio so as to make the lobes discernible
and measurable with precision, all the while protecting the CCD camera
from damage. Using astigmatic transformation, we show that the intensity
of the beam is distributed over a large area and the beam's acceleration
coefficient is deduced from the asymptotic angle (as defined by a
hyperbolic curve in elliptic coordinates). 

\begin{figure}
\begin{centering}
\includegraphics[clip,width=1\columnwidth]{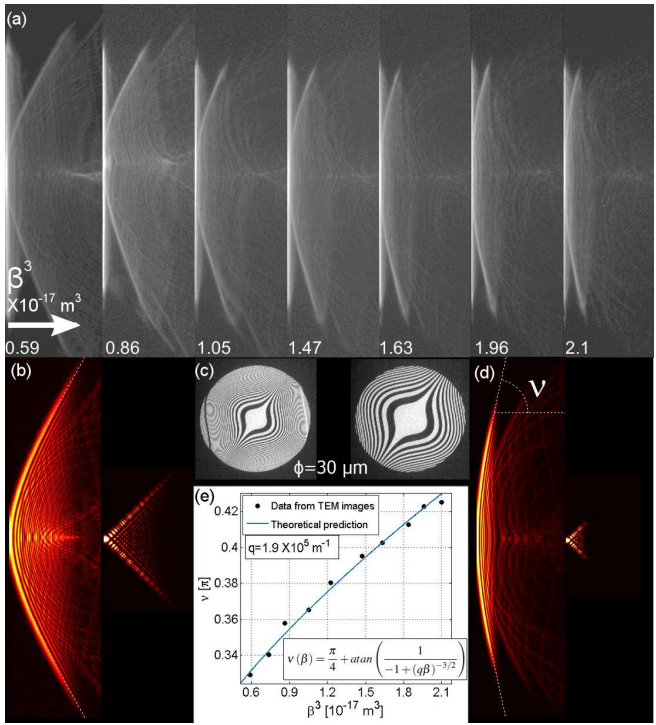}
\par\end{centering}

\centering{}\caption{\label{fig:SIMCITY_Airy} (Color online) Astigmatic Airy beams. (a)
selected measurements; (b) simulated Airy beams (right) and their
corresponding astigmatic transform (left); (c) on-axis binary masks
recorded in TEM of the lowest and highest acceleration; (d) same as
(b), for higher acceleration; (e) acceleration dependence of the hyperbola's
asymptotic angle $v$ according to the asymptotic model.}
\end{figure}

In our experiment, we designed $30\mu m$-diameter binary masks using
the expression

\begin{equation}
h\left(\xi,\eta\right)=\mathrm{sign}\left\{ \cos\left[\left(\xi^{3}+\eta^{3}\right)/3\beta^{3}\right]\right\} \label{eq:Airy-mask-1-1}
\end{equation}

where $\beta^{3}$ determines the amount of cubic phase, thereby controlling
the generated Airy's nodal trajectory. A large ensemble of these $\beta$-varying
astigmatic Airy beams were simulated, and in the resulting elliptic
coordinate-system, we naturally fitted hyperbolae to these curves,
see Fig. \ref{fig:SIMCITY_Airy}. Hyperbolae may be represented in
the positive $x$ half-space of a Cartesian coordinate system by using
the following relations:
\begin{equation}
\begin{array}{c}
x=b\cosh\left(u\right)\cos\left(v\right)\;\:\\
y=\pm b\sinh\left(u\right)\sin\left(v\right)
\end{array}
\end{equation}

where $b$ is a real positive constant, $u\in\left[0,\infty\right)$
is the curved coordinate along the hyperbola and $v\in\left[0,\pi/2\right)$
is the asymptotic angle. An animation of a simulation of the experiment
using beam-propagation methods is available on-line in the supplementary
material.

In order to tackle this problem mathematically, we perform an asymptotic
analysis of the integral (\ref{eq:eq4_F(k_xi,k_eta,a)}) using the
stationary-phase approximation. Careful inspection of the exponential's
argument yields 16 such stationary points, however, to determine the
asymptotic angle we need only to focus in the limit $x\rightarrow+\infty,\: y\rightarrow-\infty$,
where the astigmatic Airy's tail is evident and the angle $v$ is
measured. Under these conditions, it may be shown that only two stationary
points contribute to the integral. Furthermore, due to the super-exponential
decay of the Airy function in the positive $+x\rightarrow\infty$
direction, the asymptotic approximation up to terms of order $\sim a$
may be concluded to be

\begin{flushleft}
\begin{align}
 & F\left(k_{\xi},k_{\eta},\beta,a\right)\approx\frac{\pi\beta^{3/2}}{\left(-k_{\xi}k_{\eta}\right)^{1/4}}\nonumber \\
 & \times\exp\left(-\frac{2}{3}\beta^{3/2}k_{\xi}^{3/2}+2a\beta^{3}\sqrt{-k_{\xi}k_{\eta}}-\frac{2}{3}i\beta^{3/2}k_{\eta}\sqrt{-k_{\eta}}-\frac{i\pi}{4}\right).
\end{align}

\par\end{flushleft}

Now, to determine the asymptotic angle from this expression, we need
only to differentiate the real part of the exponential's argument
and find the slope. This finally yields the main result,

\begin{equation}
\nu\left(\beta\right)=\frac{\pi}{4}+\mathrm{atan}\left(\frac{1}{-1+\left(q\beta\right)^{-3/2}}\right)\label{eq:nu(beta)}
\end{equation}

where $q$ is a fitting parameter dependent, in experiment, on the
preset parameters of the optics system. The validity condition for
these approximations is $\left(q\beta\right)^{3}\ll1$, and we note
that in this elliptic system, $v\in\left(\pi/4,\pi/2\right)$. A detailed
derivation is available in the supplementary material. 

So far we have found a relationship between $\beta$, which solely
defines the mask in the fabrication, to the asymptotic angle $\nu$
measurable in experiment. We recall that in the TEM, the diffraction
plane is imaged to the CCD camera plane with magnification $M$, so
it may finally be observed. In the experiment, we maintain a collimated
input beam so the diffraction plane coincides with the focal length
$f$ of the system. It is important to understand that due to the
nonlinear trajectory of the Airy beam, the magnification factor $M$
affects the longitudinal and transverse length scales differently,
effectively reducing the observed acceleration, $\mathfrak{a}_{cam}$,
by a factor of $M$ relative to the intrinsic acceleration $\mathfrak{a}$,
so $\mathfrak{a}_{cam}=\mathfrak{a}/M$ (see supplementary).

Utilizing the theoretical background developed here and backed by
numerical simulations, we measured the asymptotic angles of 10 Airy
beams with different nodal trajectories, some of which are depicted
in Fig. \ref{fig:SIMCITY_Airy}. The same Au coated SiN membrane as
in the case of the vortex masks was used, where Airy binary gratings
(Fig.\ref{fig:SIMCITY_Airy}c) were fabricated with varying $\beta^{3}$
values in the range $\mathbf{\mathrm{\left(5.9\leq\beta^{3}\leq21\right)\unit[\times10^{-17}]{m^{3}}}}$,
which for our setup corresponded to accelerations in the range\textbf{
$\unit[\mathbf{\left(\mathrm{1.2\leq\mathfrak{a}_{cam}\leq4.1}\right)}]{\mu m^{-1}}$},
assuming $f=\unit[10]{cm}$ and measuring the magnification, $M=1015$,
directly using a Bragg diffraction grating of $400nm$ period. The
Airy gratings impose a cubic phase modulation, due to Eq.(\ref{eq:Airy-mask-1-1}),
on the beam, yielding an Airy pattern in the diffraction plane (Fig.\ref{fig:SIMCITY_Airy}b,d).
Upon enforcing elliptical astigmatism using the TEM's stigmator lenses,
the Airy beams were transformed into curved shapes, whose envelopes
were well represented by hyperbolae (experimentally in Fig.\ref{fig:SIMCITY_Airy}a,
also in simulation in \ref{fig:SIMCITY_Airy}b,d). In order to achieve
the largest angles in the vicinity of $\pi/2$, we harnessed both
the condenser and objective stigmators, carefully aligning them to
act in the same direction. Deviations from a symmetric hyperbola may
be attributed to small misalignment of the stigmators. During the
fitting process, the hyperbolic amplitude $b$ was taken as constant;
if a higher range of accelerations is desired, then the variation
of $b$ may have to be taken into account. The effect of changing
the measure of ellipticity ($a$), which is a constituent of $q$
and controlled by the stigmators, may prove useful in selection of
a different working point on the curve $\nu\left(\beta\right)$, thus
shifting or expanding the range of angles. $q$ for given system conditions
may be determined by a calibration sample. Thus, as evident from the
fit of the experimental results to Eq.(\ref{eq:nu(beta)}), shown
in Fig.\ref{fig:SIMCITY_Airy}e, we provide a method to finding the
the nodal trajectory, or acceleration coefficient, $\beta$, from
measurements of the asymptotic angle.

In this letter we examined a method of measuring the OAM of electron
vortex beams by means of elliptical transformation, previously employed
in light-optics using cylindrical lenses. The transformation is applied
literally by the turn of a knob in any standard TEM by correctly manipulating
the lens stigmators. Since OAM can be transferred between the electron
beam and the internal electron states in an atom, this method can
be readily applied for microscopic studies of materials \cite{Lloyd2012}
in electron microscopy, assuming pure-states emerge; a different approach
must be taken otherwise. We extended the scope of this transformation
and applied it to Airy beams in numerical simulation, beam-propagation
simulation and experiment, and found an analytic, asymptotic approximation
relation between the curvature of the resulting astigmatic shape and
the Airy beam's acceleration, or nodal trajectory. Our results warrant
further theoretical investigation into the astigmatic transform of
other beams to which it may be relevant, such as Bessel beams \cite{Grillo2014},
parabolic beams \cite{Bandres2008} and beams with arbitrary caustic
curves \cite{Greenfield2011,Froehly2011}, unveiling their underlying
propagation parameters.
\begin{acknowledgments}
The authors would like to acknowledge Dr. Yigal Lilach for his support
in the fabrication process. The work was supported by the Israel Science
Foundation, grant no. 1310/13, and the Australian Research Council.
\end{acknowledgments}
\bibliographystyle{apsrev}
\addcontentsline{toc}{section}{\refname}\bibliography{references}

\end{document}


\title{Supplementary Material - Unveiling the orbital angular momentum and
acceleration of electron beams}

\author{Roy Shiloh}

\email{royshilo@post.tau.ac.il}

\affiliation{Department of Physical Electronics, Fleischman Faculty of Engineering,
Tel Aviv University, Tel Aviv 6997801, Israel}

\author{Yuval Tsur}

\affiliation{Department of Physical Electronics, Fleischman Faculty of Engineering,
Tel Aviv University, Tel Aviv 6997801, Israel}

\author{Yossi Lereah}

\affiliation{Department of Physical Electronics, Fleischman Faculty of Engineering,
Tel Aviv University, Tel Aviv 6997801, Israel}

\author{Boris A. Malomed}

\affiliation{Department of Physical Electronics, Fleischman Faculty of Engineering,
Tel Aviv University, Tel Aviv 6997801, Israel}

\author{Vladlen Shvedov}

\affiliation{Laser Physics Centre, The Australian National University, Canberra
ACT 0200, Australia}

\author{Cyril Hnatovsky}

\affiliation{Laser Physics Centre, The Australian National University, Canberra
ACT 0200, Australia}

\author{Wieslaw Krolikowski}

\affiliation{Laser Physics Centre, The Australian National University, Canberra
ACT 0200, Australia}

\author{Ady Arie}

\affiliation{Department of Physical Electronics, Fleischman Faculty of Engineering,
Tel Aviv University, Tel Aviv 6997801, Israel}

\maketitle

\section{Measurements of OAM of $10\hbar$}

\begin{figure}[H]
\begin{centering}
\includegraphics{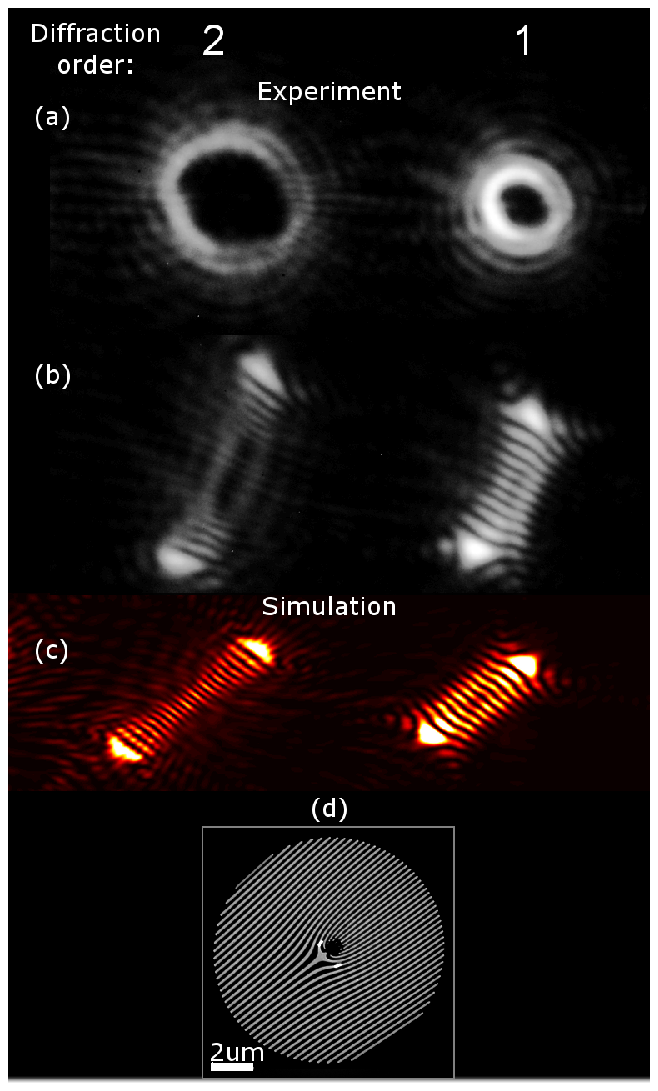}
\par\end{centering}

\centering{}\caption{\label{fig:charge10} (Color online) Experimental results, showing
two diffraction orders in focus: (a) vortices generated by a charge
10 fork grating, using a focused stigmatic beam; (b) corresponding
vortices after elliptic transformation; (c) Simulation results. (d)
TEM images of the charge 10 fork ($\unit[0.5]{\mu m}$ period). The
two bright spots near the center of the fork are fabrication errors.}
\end{figure}

We repeated the experiment for a charge 10 vortex beam, generated
by a fork grating with a $\unit[0.5]{\mu m}$ carrier period: as seen
in Fig. \ref{fig:charge10}. A numerical beam-propagation simulation
(Fig. \ref{fig:charge10}c) shows similar results, where 10 dark stripes
may be counted in the first order in both simulation and experiment.
The second order counts 20 stripes, however, it is harder to distinguish
between them due to technical limits of the CCD and stability of the
TEM. It is also worth mentioning that, if the astigmatic transformation
is not completely elliptic, higher orders will only be partially converted.
This may be observed in the dark patch overlaying the second order
in the experiment (Fig. \ref{fig:charge10}b), while the ``braided''
dark fringes are a result of overlap of the next high-order mode.
We can therefore conclude that the elliptic coordinate transformation
method enables to measure vortex beam with an OAM of at least $10\hbar$.

\section{Numerical inspection of the astigmatic Airy integral}

We numerically evaluate the two-dimensional integral,

\begin{equation}
F\left(x,y,\beta,a\right)=\int_{-\infty}^{\infty}\int_{-\infty}^{\infty}\exp\left[i\phi\left(\xi,\eta\right)\right]d\xi d\eta\label{Main_integral}
\end{equation}

with the phase

\begin{equation}
\phi\left(\xi,\eta\right)=x\xi+y\eta+2a\xi\eta+\left(3\beta^{3}\right)^{-1}\left(\xi^{3}+\eta^{3}\right).\label{eq:phi}
\end{equation}

In the Letter, we use the notation $k_{\xi}\equiv x$ and $k_{\eta}\equiv y$
to signify that the result is in the diffraction plane. Here we use
$x,y$ for elegance of the mathematical derivation.

\begin{figure}[h]
\begin{centering}
\includegraphics[clip]{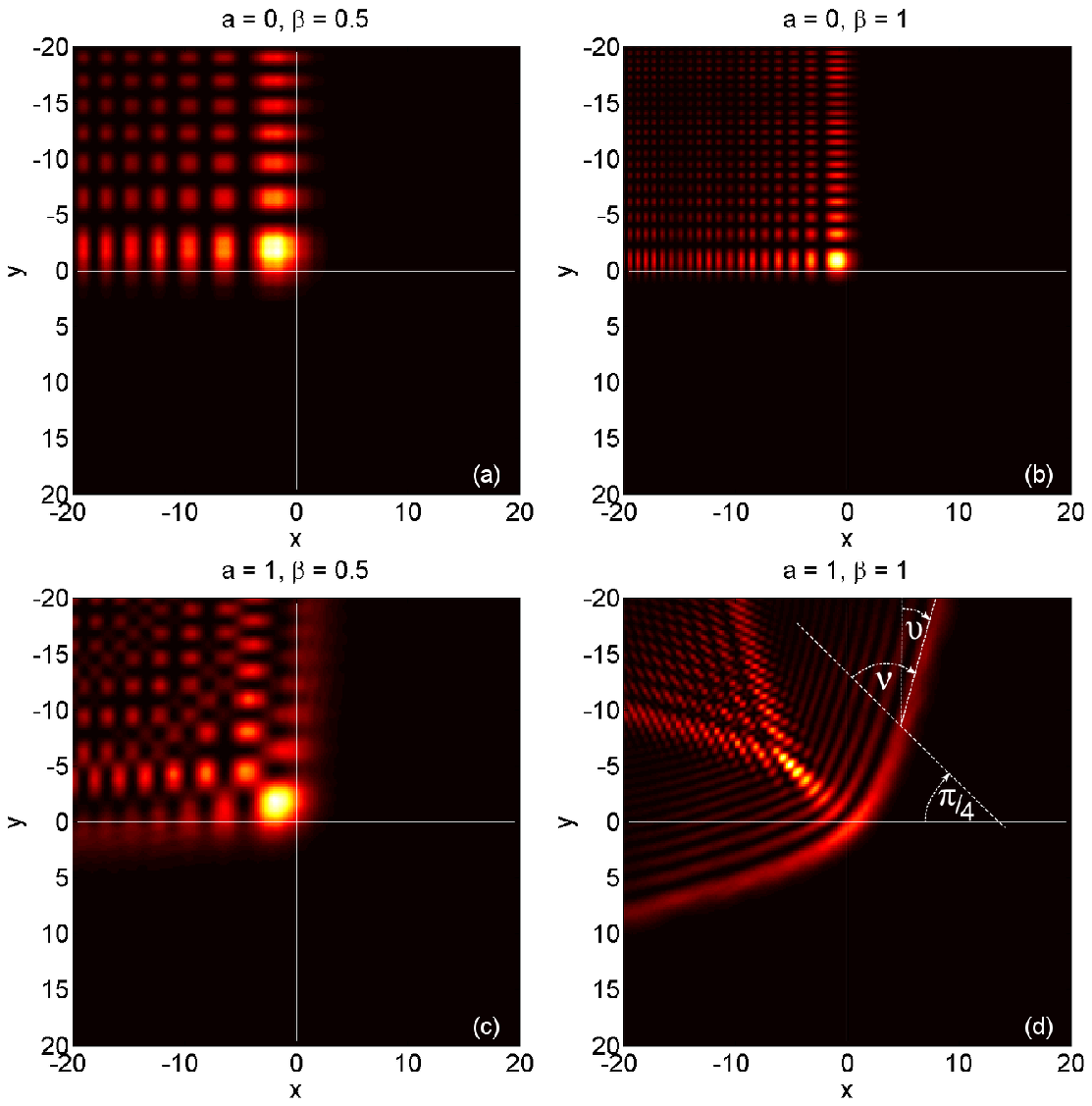}
\par\end{centering}

\centering{}\caption{\label{fig:numericalAiry} Numerical evaluation of expression (\ref{Main_integral}):
(a,b) stigmatic ($a=0$) Airy and (c,d) astigmatic ($a=1$) Airy for
$\beta=0.5,1$. Definitions of the angles are depicted in (d).}
\end{figure}

It is physically acceptable to interpret $\beta$ as the parameter
defining the nodal trajectory curve of the Airy beam. The parameter
$a$ is a measure of the efficacy of the elliptic transformation.
For the two example cases depicted in Fig.\ref{fig:numericalAiry},
we see that for the stigmatic Airy ($a=0$, Fig.\ref{fig:numericalAiry}a,b),
the asymptotic angle $\upsilon$ equals $0$, measured from the negative
y-axis to a vertical line well-defined by the Airy lobes, regardless
of $\beta$. For the astigmatic Airy ($a=1$, Fig.\ref{fig:numericalAiry}c,d),
$\beta$ affects the result of the integral by changing the asymptotic
angle $\upsilon$: as the parameter $\beta$ increases, so does the
angle $\upsilon$, measured from from the negative y-axis to the asymptotic
tail of the transformed two-dimensional Airy pattern (see Fig.\ref{fig:numericalAiry}d).
To recover the asymptotic angle analytically, we must focus on the
region $+x>0$, $-y>0$, and specifically $+x\rightarrow\infty$,
$-y\rightarrow\infty$.

\section{Asymptotic approximation of the astigmatic Airy integral}

Here we solve the integral (\ref{Main_integral}) by asymptotic means.
We first note an obvious symmetry,

\begin{equation}
F\left(x,y,\beta,a\right)=F^{\ast}\left(x,y,\beta,-a\right)
\end{equation}

where $\ast$ stands for the complex conjugate. Thus, $F$ is purely
real only at $a=0$. It is also noted that for $a=0$ the result of
the integral is proportional to the two-dimensional Airy function,
$\mbox{Ai}\left(x\right)\mbox{Ai}\left(y\right)$.

For the asymptotic evaluation of the integral, it is necessary to
find the stationary-phase point (SPP), $\left(\xi_{0},\eta_{0}\right)$,
at which $\phi_{\xi}=\phi_{\eta}=0$ (the subscripts stand for partial
derivatives). As follows from (\ref{eq:phi}), the equations which
determine the SPP are
\begin{eqnarray}
x+2a\eta_{0}+\beta^{-3}\xi_{0}^{2} & = & 0,\nonumber \\
y+2a\xi_{0}+\beta^{-3}\eta_{0}^{2} & = & 0.\label{eq:quadratic_coupled_SP}
\end{eqnarray}

These two coupled quadratic equations in $\xi_{0},\eta_{0}$ may be
decoupled into two quartic equations yielding 16 roots, which are
the SPPs. However, in the limits $+x\rightarrow\infty$, $-y\rightarrow\infty$,
the expression is oscillating along $y$ and evanescent along $x$;
then, the SPPs may be determined by asymptotic expansion to leading
order in $a$, yielding two relevant sets of coordinates,
\begin{gather}
\xi_{0}\approx i\beta^{3/2}\sqrt{x}\pm ia\beta^{3}\sqrt{-\frac{y}{x}},\nonumber \\
\eta_{0}\approx\pm\beta^{3/2}\sqrt{-y}\mp ia\beta^{3}\sqrt{-\frac{x}{y}},\label{xi02}
\end{gather}
as indicated with the help of $\pm$. At these two SPPs, the expression
for the phase, taken in the present approximation, is written
\begin{equation}
i\phi\left(\xi_{0},\eta_{0}\right)\approx-\frac{2}{3}\beta^{3/2}x^{3/2}\pm\frac{2}{3}i\beta^{3/2}y\sqrt{-y}\mp2a\beta^{3}\sqrt{-xy}.\label{phi02}
\end{equation}
The second derivatives at the SPPs are:
\begin{equation}
\phi_{\xi\xi}\left(\xi_{0},\eta_{0}\right)\approx2i\beta^{-3/2}\sqrt{x},\:\phi_{\eta\eta}\left(\xi_{0},\eta_{0}\right)\approx\pm2\beta^{-3/2}\sqrt{-y},\:\phi_{\xi\eta}\left(\xi_{0},\eta_{0}\right)=2a.\label{second2}
\end{equation}
Finally, the computation of the integral (\ref{Main_integral}) in
the asymptotic approximation yields the following result:
\begin{gather}
F\left(x,y,\beta,a\right)\approx\frac{\pi\beta^{3/2}}{\left(-xy\right)^{1/4}}\exp\left(-\frac{2}{3}\beta^{3/2}x^{3/2}\right)\nonumber \\
\times\left[\exp\left(-2a\beta^{3}\sqrt{-xy}+\frac{2}{3}i\beta^{3/2}y\sqrt{-y}+\frac{i\pi}{4}\right)+\exp\left(2a\beta^{3}\sqrt{-xy}-\frac{2}{3}i\beta^{3/2}y\sqrt{-y}-\frac{i\pi}{4}\right)\right]\label{apprappr2}
\end{gather}

We check this approximation at $a=0$ and note that it indeed reduces
to the product of asymptotic approximations for the usual Airy functions,
in which case expression (\ref{apprappr2}) is also purely real.

However, at $a\neq0$ it is complex. In fact, in the case of $a>0$
and $-xy\rightarrow\infty$, the second term in the square brackets
is exponentially large in comparison to the first term, hence (\ref{apprappr2})
may be reduced to the one with a single leading term, 
\begin{equation}
F\left(x,y,\beta,a\right)\approx\frac{\pi\beta^{3/2}}{\left(-xy\right)^{1/4}}\exp\left(-\frac{2}{3}\beta^{3/2}x^{3/2}+2a\beta^{3}\sqrt{-xy}-\frac{2}{3}i\beta^{3/2}y\sqrt{-y}-\frac{i\pi}{4}\right).\label{apprapprappr}
\end{equation}

Formally, both expressions (\ref{apprappr2}) and (\ref{apprapprappr})
may give an exponentially large result in the case of $a^{2}\beta^{3}\left(-y\right)\gg x^{2}$.
However, the asymptotic approximation is not valid in this case, as
its validity is determined by the condition that corrections $\sim a$
in Eqs.(\ref{xi02}) are small in comparison with the leading terms,
i.e., $a^{2}\beta^{3}\left(-y\right)\ll x^{2}$, $a^{2}\beta^{3}x\ll y^{2}$.

\section{Mathematical derivation of the relation $\nu\left(\beta\right)$}

In the limit $+x\rightarrow\infty$, $-y\rightarrow\infty$, the expression
in Eq.\ref{apprapprappr} produces linear, parallel contours with
an angle to the axes. Assuming $+x,-y$ tend to infinity equally fast,
i.e. $+x=-y\triangleq\tilde{q}\rightarrow\infty$, this angle is solely
determined by the real part of the exponent's argument of Eq.(\ref{apprapprappr}),

\begin{equation}
\varphi\left(x,y,\beta,a\right)=-\frac{2}{3}\beta^{3/2}x^{3/2}+2a\beta^{3}\sqrt{-xy},
\end{equation}

by calculating the slope,

\begin{equation}
Slope=\frac{\varphi_{y}}{\varphi_{x}}=\frac{ax}{ay+x\sqrt{-y\beta^{-3}}}=\frac{1}{-1+\left(q\beta\right)^{-3/2}}
\end{equation}

where for elegance we define $q=\tilde{q}^{-1/3}a^{2/3}$ and $q>0$.
The angle of this slope is then given by $\upsilon=\mbox{atan}\left(Slope\right)$
with respect to the negative y-axis . In the experiment, however,
it is easier to determine the angle relative to the inversion symmetry
of the astigmatic Airy's curve, i.e. following a $-45{}^{\circ}$
(clockwise) rotation of the image relative to the (still horizontal
and vertical) $x,y$ axes. In this case, we define the angle $\nu=\pi/4+\upsilon$,
measured from the negative x-axis to the asymptotic tail of the astigmatic
Airy (see \ref{fig:numericalAiry}d). Since in the proposed elliptical
system $\upsilon\in(0,\pi/4)$, it is now obvious that $\nu\in(\pi/4,\pi/2)$,
owing to the fundamental right-angle appearance of the stigmatic Airy
pattern. We therefore write the final result,

\begin{equation}
\nu\left(\beta\right)=\frac{\pi}{4}+atan\left(\frac{1}{-1+\left(q\beta\right)^{-3/2}}\right)
\end{equation}

where $q$ is now a fitting parameter dependent, in experiment, on
the stigmator lens through $a$ and the effective scaling through
$\tilde{q}$. The exact dependence may be deduced by further investigation.
To finalize the discussion, recall that the asymptotic approximation
is valid, under the assumptions above, for $a^{2}\beta^{3}\ll\tilde{q}$.
Not incidentally, this may be rewritten $\left(q\beta\right)^{3}\ll1$
and in given experimental conditions may be approximated further.
An animation of a simulation of the experiment using beam-propagation
methods is available on-line in the supplementary material.

\section{Derivation of the nodal trajectory\textmd{\textup{\normalsize{ }}}}

The integral representation of the Airy function \cite{Vallee2004}
may be rewritten to yield 

\begin{equation}
\int_{-\infty}^{\infty}\exp\left[i\left(\frac{\xi^{3}}{3\beta^{3}}+\frac{zk_{DB}}{2f^{2}}\xi^{2}+k_{\xi}\xi\right)\right]d\xi=2\pi\beta\exp\left[-\frac{iz\beta^{3}k_{DB}}{2f^{2}}\left(k_{\xi}-\frac{\beta^{3}k_{DB}^{2}}{6f^{4}}z^{2}\right)\right]\text{Ai}\left[\beta\left(k_{\xi}-\frac{\beta^{3}k_{DB}^{2}}{4f^{4}}z^{2}\right)\right],\label{eq:AiryInt}
\end{equation}

where the derivation is trivially extendable to (2+1)D. Under the
paraxial approximation, the length scale may be written as

\begin{equation}
\beta k_{\xi}=\beta\frac{k_{DB}}{f}x\equiv\frac{x}{x_{0}}.\label{eq:beta_def}
\end{equation}

Thus the argument of the Airy function now gives the familiar form
\cite{Voloch-Bloch2013},

\begin{equation}
\text{Ai}\left[\frac{1}{x_{0}}\left(x-\frac{1}{4k_{DB}^{2}x_{0}^{3}}z^{2}\right)\right]
\end{equation}

and the nodal trajectory parameter, sometimes referred to as the acceleration
parameter, is understood to be

\begin{equation}
\mathfrak{a}=\frac{1}{2k_{DB}^{2}x_{0}^{3}}.\label{eq:acceleration}
\end{equation}

In the TEM, the diffraction plane is imaged to the CCD camera plane
with magnification $M$. We may therefore define the transverse $x_{cam}$
and on-axis $z_{cam}$ displacements with this magnification, 

\begin{equation}
x_{cam}=M\cdot\frac{z^{2}}{4k_{DB}^{2}x_{0}^{3}},\label{eq:xMagnification}
\end{equation}
\begin{equation}
z_{cam}=M\cdot z.\label{eq:zMagnification}
\end{equation}

It therefore follows that 
\begin{equation}
x_{cam}\left(z_{cam}\right)=\frac{1}{M}\frac{z_{cam}^{2}}{4k_{DB}^{2}x_{0}^{3}}\label{eq:x_cam_magnified}
\end{equation}

and understood that in the camera plane, the observed parameter $\mathfrak{a}_{cam}$
obeys $\mathfrak{a}_{cam}=\mathfrak{a}/M$. Thus, we finally obtain
\begin{equation}
\mathfrak{a}_{cam}=\frac{\beta^{3}k_{DB}}{2Mf^{3}}.
\end{equation}

\bibliographystyle{apsrev}
\addcontentsline{toc}{section}{\refname}\bibliography{references}